\documentclass[pdflatex,sn-nature]{sn-jnl}

\usepackage{graphicx}%
\usepackage{multirow}%
\usepackage{amsmath,amssymb,amsfonts}%
\usepackage{mathrsfs}%
\usepackage{xcolor}%
\usepackage{textcomp}%
\usepackage{manyfoot}%
\usepackage{booktabs}%
\usepackage[title]{appendix}%

\usepackage{gensymb}%
\usepackage{siunitx}%
\DeclareSIUnit{\um}{\micro\meter}%
\DeclareSIUnit{\mm}{\milli\meter}%
\DeclareSIUnit{\cm}{\centi\meter}%
\DeclareSIUnit{\m}{\meter}%
\DeclareSIUnit{\kg}{\kilo\gram}%
\DeclareSIUnit{\ns}{\nano\second}%
\DeclareSIUnit{\ps}{\pico\second}%
\DeclareSIUnit{\eV}{\electronvolt}%
\DeclareSIUnit{\keV}{\kilo\electronvolt}%
\usepackage{bm}%
\usepackage{subcaption}%
\usepackage[export]{adjustbox}%

\begin{document}

\title[Control of Magnetic Reconnection in HED Plasmas]{Control of Magnetic Reconnection in High Energy Density Plasmas}

\author*[1]{\fnm{J. L.} \sur{Latham}}\email{joshla@umich.edu}
\author[2]{\fnm{B. K.} \sur{Russell}}
\author[3,4]{\fnm{C.} \sur{Dong}}
\author[5]{\fnm{C. A.} \sur{Walsh}}
\author[6]{\fnm{K. G.} \sur{Miller}}
\author[1]{\fnm{P. T.} \sur{Campbell}}
\author[1]{\fnm{L.} \sur{Willingale}}
\author[6]{\fnm{P.} \sur{Nilson}}
\author[1]{\fnm{K.} \sur{Krushelnick}}

\affil[1]{\orgdiv{Center for Ultrafast Optical Science}, \orgname{University of Michigan}, \orgaddress{\city{Ann Arbor}, \state{Michigan}, \postcode{48109}, \country{USA}}}
\affil[2]{\orgdiv{Department of Astrophysics}, \orgname{Princeton University}, \orgaddress{\city{Princeton}, \state{NJ}, \postcode{08544}, \country{USA}}}
\affil[3]{\orgdiv{Center for Space Physics and Department of Astronomy}, \orgname{Boston University}, \orgaddress{\city{Boston}, \state{MA}, \postcode{02215}, \country{USA}}}
\affil[4]{\orgdiv{School of Natural Sciences}, \orgname{Institute for Advanced Study}, \orgaddress{\city{Princeton}, \state{NJ}, \postcode{08540}, \country{USA}}}
\affil[5]{\orgname{Lawrence Livermore National Laboratory}, \orgaddress{\street{7000 East Avenue}, \city{Livermore}, \state{California}, \postcode{94550}, \country{USA}}}
\affil[6]{\orgdiv{Laboratory for Laser Energetics}, \orgname{University of Rochester}, \orgaddress{\city{Rochester}, \state{New York}, \postcode{14623}, \country{USA}}}

\abstract{Magnetic reconnection governs the explosive release of magnetic energy in systems from the solar corona to fusion plasmas, yet controlling it in the laboratory has remained out of reach. Here we demonstrate active control of reconnection in high-power laser-driven plasmas using a third, relativistic-intensity laser pulse that injects filaments of electron current into the reconnecting system. Two moderate-intensity lasers drive colliding magnetized plumes that reconnect, forming plasmoids in the current sheet as seen in proton deflectometry. The relativistic laser generates magnetic fields matching the polarity on either side of the layer, and, depending on its arrival time, either accelerates the breakup of the current sheet or suppresses reconnection. Arriving early, before the plumes strongly interact, it builds a pocket of magnetic pressure that repels them via flux pileup; arriving after the current sheet forms, it accelerates electrons that extend current filamentation instabilities into the upstream, causing rapid dissipation of the reconnecting magnetic field. This approach opens a route to steering magnetic energy flow in fusion plasmas and broadens the range of systems accessible to laboratory astrophysics.}

\maketitle
Magnetic reconnection is the process by which oppositely directed magnetic fields break and reconnect, suddenly converting stored magnetic energy into kinetic energy and energetic particles \cite{yamadaMagneticReconnection2010,jiMagneticReconnectionEra2022,totoricaNonthermalElectronEnergization2016}.
It is a leading candidate for explaining exotic astrophysical phenomena such as solar flares \cite{masudaLooptopHardXray1994}, gamma-ray bursts \cite{medvedevGenerationMagneticFields1999,lyubarskyAreGammaRayBurst2006,uzdenskyMagneticReconnectionExtreme2011}, and X-ray emission from pulsar wind nebulae \cite{ceruttiModellingHighenergyPulsar2016}.
It occurs where two plasmas carrying oppositely directed magnetic fields meet, forming a sheet of electric current at the collision interface; as the field lines reorganize into a lower-energy configuration, the released energy drives plasma flows along the original field direction, transverse to the inflow \cite{sweet1958IAUS6123S1958,parkerSweetsMechanismMerging1957}.
When the current sheet becomes sufficiently elongated, it is unstable to tearing into multiple magnetic islands, or ``plasmoids'' (or flux ropes in 3D), which form in both collisional and collisionless reconnection and strongly shape the reconnection dynamics and particle energization \cite{loureiroInstabilityCurrentSheets2007,bhattacharjeeFastReconnectionHighLundquistnumber2009,daughtonTransitionCollisionalKinetic2009,foxNovelKineticMechanism2021}.
However, if magnetic reorganization at the collision interface happens slower than the incoming advection of magnetic field, then flux pileup occurs \cite{foxFastMagneticReconnection2011,rosenbergLaboratoryStudyAsymmetric2015}, enhancing the reconnection rate for dissipative processes that depend on the current density but also increasing the magnetic pressure.  

Until now, experimental studies of magnetic reconnection in high-energy-density (HED) plasmas have been limited to plasma formation and magnetic field generation on a single timescale.  
With some exceptions \cite{raymondRelativisticelectrondrivenMagneticReconnection2018,palmerFieldReconstructionProton2019,lawRelativisticMagneticReconnection2020}, these studies have been on a non-relativistic, nanosecond timescale, with micrometer to millimeter length scales, using high-power lasers \cite{nilsonMagneticReconnectionPlasma2006,liObservationMegagaussFieldTopology2007,willingaleProtonDeflectometryMagnetic2010,rosenbergSlowingMagneticReconnection2015,tubmanObservationsPressureAnisotropy2021} or pulsed power \cite{hareAnomalousHeatingPlasmoid2017} to generate the plasma and the magnetic fields.  
In these experiments, reconnection is observed between two colliding hemispherical or cylindrical plasmas with magnetic field that is either internally generated through the Biermann battery effect \cite{nilsonMagneticReconnectionPlasma2006,liObservationMegagaussFieldTopology2007,willingaleProtonDeflectometryMagnetic2010,tubmanObservationsPressureAnisotropy2021} or externally generated through an external electric current \cite{fikselMagneticReconnectionColliding2014,kuramitsuMagneticReconnectionDriven2018,chienStudyMagneticallyDriven2019}.
The conditions at the current sheet are determined by the pressure and temperature that drive the inflow of plasma, and plasmoid formation, if observed \cite{hareAnomalousHeatingPlasmoid2017,pearcyExperimentalEvidencePlasmoids2024}, is determined by fluctuations.  
These experiments have proven useful for understanding the magnetic dynamics and particle acceleration in plasmas relevant to astrophysics and inertial fusion energy, 
but due to the self-generated nature of the current sheet, 
there is still a limited understanding of the role of modulated electron velocity distribution in the reconnection dynamics under HED conditions.

In this paper, we demonstrate the control of reconnection in HED plasmas by an external perturbation in the form of a short-pulse (SP), relativistic-intensity laser impinging directly upon the current sheet. 
The relativistic laser accelerates a beam of electrons whose current runs parallel to the current in the sheet. This beam generates a column of magnetic field with the same polarity as the field on either side of the current sheet.
If the relativistic laser arrives after the current sheet has been fully established, then it seeds additional magnetic structures in the sheet and in the upstream that drive rapid breakup and dissipation of the entire 3D electric current that feeds the sheet.  
However, if the relativistic laser arrives when the current sheet is just beginning to form, then the magnetic pressure created by the electron beam magnetic field is compressed by the inflowing magnetic field from the plasma plumes, 
so that the increased magnetic pressure opposes the inflow of plasma and prevents formation of a reconnection current sheet.
The control of reconnection using this or a similar technique could be further refined to enable on-demand breakup of a current sheet or tuning of its structure, opening up new possibilities in reconnection studies and in the control of fusion plasmas. 

\begin{figure}
    \centering
    \includegraphics[width=\linewidth]{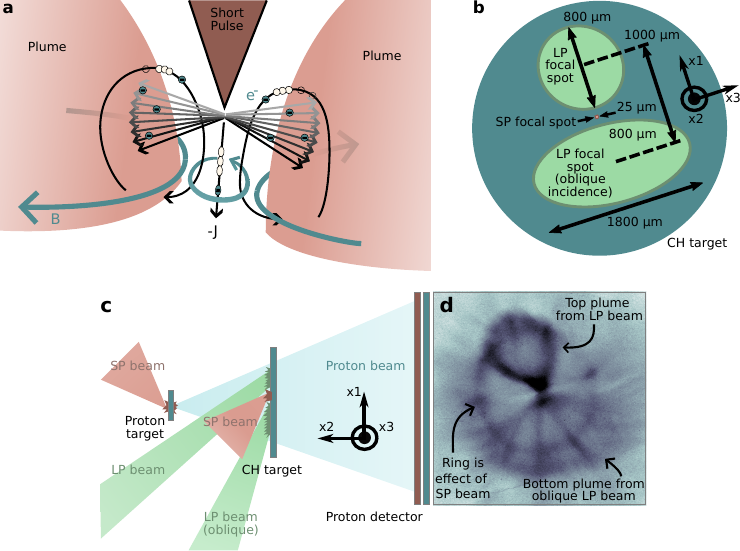}
    \caption{
    \textbf{Experiment setup for laser-controlled magnetic reconnection.}
    \textbf{a} Schematic of experimental concept: a short-pulse relativistic laser (``Short Pulse'') changes the topology of the field lines ($\bm{B}$) by injecting a negative current ($-\bm{J}$) of electrons ($e^-$) into the current sheet and radially accelerating electrons into the upstream plasma, inducing current instabilities. 
    \textbf{b} Schematic of long-pulse (LP) laser focal spots and short-pulse (SP) laser focal spot on the main plastic (CH) target.
    \textbf{c} Schematic of setup for experiment, not drawn to scale. Proton target is Cu foil with Ta shield. CH target is \SI{50}{\micro\meter} thick polystyrene. Proton detector is radiochromic film stack with Al shielding.
    \textbf{d} Example proton deflectometry image on radiochromic film (RCF).}
    \label{fig:setup}
\end{figure}

\section*{Results}

\subsection*{Experimental results}

\begin{figure}
    \centering
    \includegraphics[center,width=\linewidth]{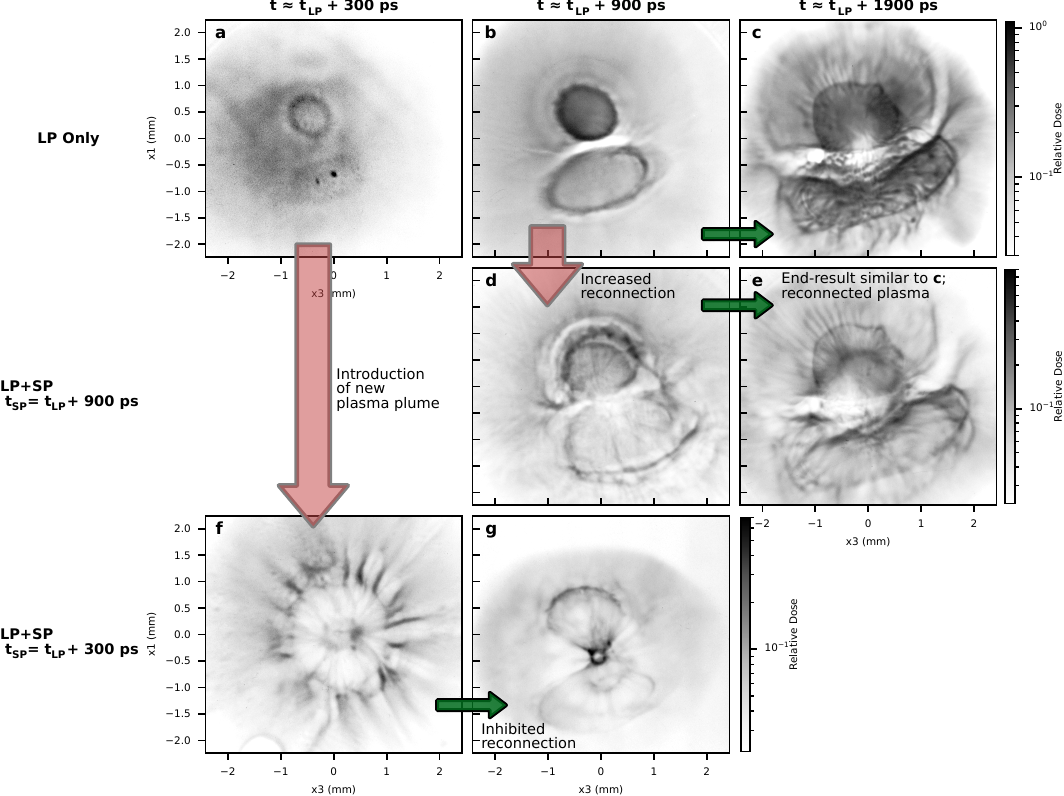}
    \caption{\label{fig:doryboard} 
    \textbf{Control of reconnection by a relativistic laser.}
    Axes are scaled to the object plane.
    Panels \textbf{a-c} show the development of the ultraviolet long-pulse (LP) driven plumes without any short-pulse (SP). \textbf{d-e} show the case of firing the SP at $t=t_\text{LP}+900\text{ ps}\pm25\text{ ps}$, and \textbf{f-g} show the case of firing the SP at $t=t_\text{LP}+300\text{ ps}\pm 25\text{ ps}$.  
    Each column of deflectometry images represents approximately the same amount of total LP drive at the time of the deflectometry image. 
    }
\end{figure}

Two long-pulse (LP) lasers irradiated the surface of a plastic foil to generate plumes of plasma, and an SP irradiated the plumes at their collision interface after they had collided (Fig.\ \ref{fig:setup}).
Each of the LP-generated plumes self-magnetizes due to the Biermann battery effect \cite{schlueterUeberUrsprungMagnetfelder1950,stamperSpontaneousMagneticFields1971}, so that when they collide, the oppositely-magnetized plasma from the outer edge of each plume flows in, forming a sheet of electric current, and magnetic reconnection ensues (Fig.\ \ref{fig:doryboard}\textbf{a}-\textbf{c}). 
The relativistic laser pulse, having a 10 ps full-width at half-maximum (FWHM) duration, is 250 times shorter than the laser pulses used to drive the plasma plumes and thus imposes a sudden change in the plasma conditions at and around the current sheet during the reconnection process.  

\begin{figure}
\centering
\includegraphics[width=\linewidth]{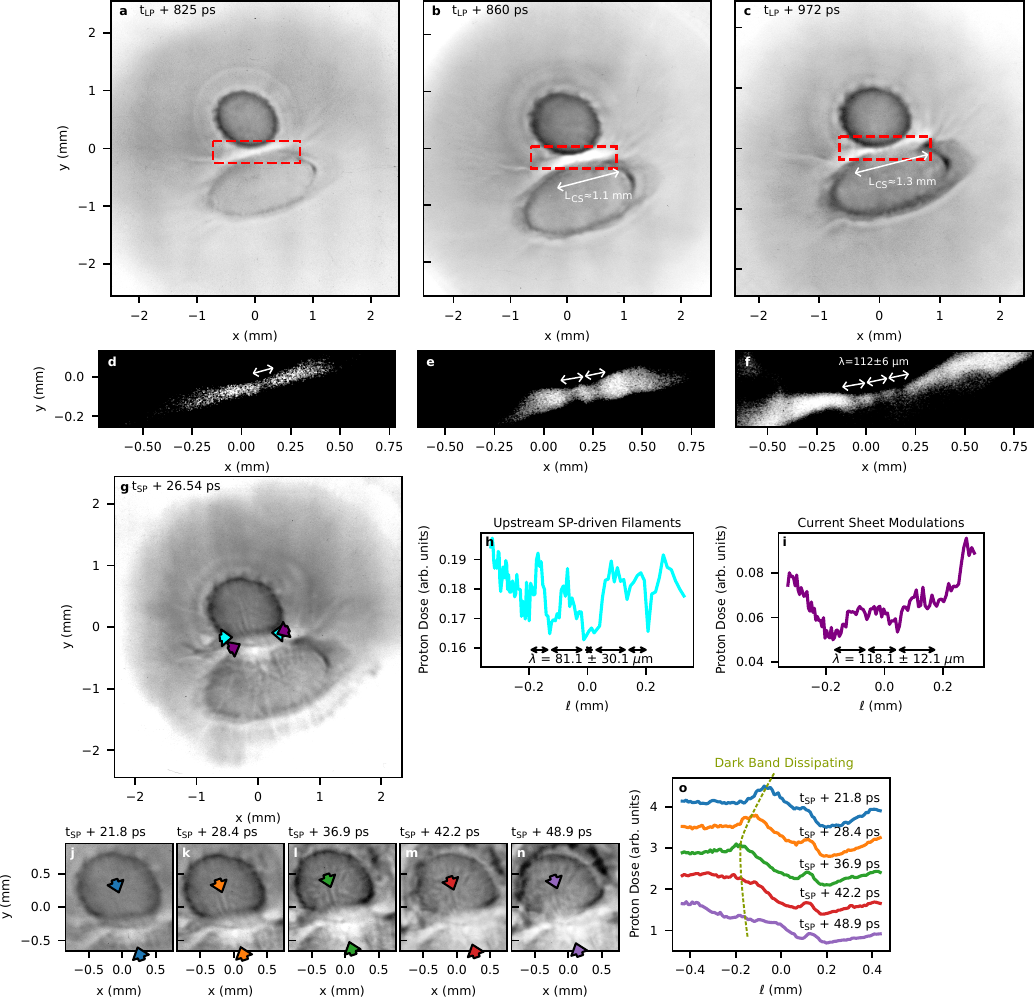}
  \caption{
    \textbf{Long-pulse current sheet plasmoids and short-pulse laser-driven dissipation.}
    Panels \textbf{a-f} are long-pulse-only proton deflectometry images.
    Panels \textbf{d-f} are high-contrast close-ups of the regions in the red dashed boxes in panels \textbf{a-c}, respectively, to make the modulations in the current sheet (CS) more easily visible.
    Panel \textbf{g} shows a proton deflectometry image at $t_\text{SP}+ 26.54$ ps ($t_\text{SP}$ is uncertain to $\pm 25$ ps), 900 ps after the onset of the LP beams.
    The colored arrows in \textbf{g} denote the boundaries of the lineouts shown in \textbf{h} and \textbf{i}.
    The cyan/lighter lineout is of the dark band at the bottom of the top plume, and the purple/darker lineout is in the current sheet.
    The black double-arrows in \textbf{h}
    and \textbf{i} 
    mark the trough-to-trough wavelengths of modulations, and the mean and standard deviation wavelength is annotated.
    \textbf{j}-\textbf{n} are 5 images from the same film stack with peak proton energy deposition corresponding to a time difference of about 7 ps between images, highlighting the motion of the dark band at the bottom of the top plume.
  The lineouts between the colored arrows in \textbf{j}-\textbf{n} are shown in \textbf{o}, with the location of the peak of the dark band marked by the overlaid dashed line.}
  \label{fig:37102_superlineout}
\end{figure}

The reconnecting current sheet between the LP-generated plasma plumes is unstable and may break up into a series of plasmoids (Fig.\ \ref{fig:37102_superlineout} \textbf{a}-\textbf{f}). 
Plasmoid formation has been observed in reconnecting current sheets at these scales under similar experimental conditions \cite{pearcyExperimentalEvidencePlasmoids2024}.
After 10 estimated growth times $\gamma^{-1}$ of the plasmoid instability (Supplementary Sec.\ \ref{sec:lponlyplasmoid}), a stochastic array of plasmoids develops as the plumes continue to reconnect (Fig.\ \ref{fig:doryboard}\textbf{c}).  
The LP lasers continually irradiated the target throughout all measurements of the magnetic field, therefore at late times the LP-only reconnection erupts into a myriad of instabilities rather than stagnating  (cf.\ \cite{rosenbergSlowingMagneticReconnection2015}).

When the SP laser was fired at the plasma plume collision interface during the initial onset of spontaneous plasmoid formation, it accelerated hot electrons toward the target, driving a sudden development of current structures in the current sheet and upstream and dissipating the reconnecting magnetic field. 
Proton deflectometry measurements reveal that the area affected by the relativistic laser was dominated by counterclockwise magnetic fields (Fig.\ \ref{fig:doryboard}\textbf{d}, Supplementary Fig.\ \ref{fig:37102bfieldinversion}). 
This polarity of magnetic field is associated with a current of electrons flowing in toward the target from the front surface direction.  
The extended underdense plasma, estimated to have a density scale length of \SI{125}{\um} to \SI{250}{\um}, facilitates the relativistic laser acceleration of electrons in the laser propagation direction, which is 45\textdegree\ to target normal, via the direct laser acceleration (DLA) or $\bm{J}\times\bm{B}$ mechanism \cite{pukhovRelativisticMagneticSelfChanneling1996}. 
After the initial electron acceleration which occurs over the duration of the 10 ps laser pulse, the accelerated electrons cause multiple current filaments to form (Fig.\ \ref{fig:37102_superlineout}\textbf{g}-\textbf{h}) with thickness on the order of the ion inertial length (Supplementary Sec.\ \ref{app:params}), on top of the existing plasmoid structure (Fig.\ \ref{fig:37102_superlineout}\textbf{a}-\textbf{f}, Fig.\ \ref{fig:37102_superlineout}\textbf{i}), causing dissipation on a tens of ps timescale (Fig.\ \ref{fig:37102_superlineout}\textbf{j}-\textbf{o}), which is about an order of magnitude faster than the dissipation observed in the LP-only case. 
As the LPs continue to drive the plasma, reconnection continues to develop, resulting in a similar multitude of plasmoids as in the unperturbed case, but in a slightly more dissipated state (Fig.\ \ref{fig:doryboard}, \textbf{e} vs. \textbf{c}).

The SP has the opposite effect on the reconnection of the two plumes if it arrives between the plumes prior to the development of the current sheet
because the magnetic field it generates is compressed against the plume magnetic field on either side of the collision region, building up magnetic pressure that opposes the inflow of plasma.
The initial magnetic field created by the SP has a clockwise component that spreads over the outer surface of the LP-driven plumes, creating a circular feature in proton deflectometry images that expands over a tens of picosecond timescale (Fig.\ \ref{fig:doryboard} \textbf{f}), 
generated by the expanding hot electrons.
However, there is an interior region of magnetic field between the two LP-driven plumes with a counterclockwise magnetic polarity.
This polarity of magnetic field is associated with the beam of electrons accelerated in the underdense plasma by the relativistic laser.
The small circle of beam-driven magnetic field is such that its directionality matches that of the LP-driven plume's field on either side of the collision region.
As the LP-driven plumes expand, they compress the fields initially generated by the SP and are pushed back by these fields,
instead of entering a reconnection geometry.  
The impeding counterclockwise field is still visible in deflectometry images as a dark ring at the point in time when the reconnection current would otherwise have started developing plasmoids (Fig.\ \ref{fig:doryboard} \textbf{g}). 
Through an increased magnetic pressure from same-polarity magnetic fields, the laser-driven magnetic field therefore inhibits reconnection if it arrives before the LP-driven plumes have collided and formed a current sheet. 

In both cases the relativistic laser drives an electron current toward the target that generates a counterclockwise magnetic field; the opposite effects on reconnection arise from whether this perturbation is injected into an already-formed current sheet or ahead of it. When injected into an established sheet, the beam current runs parallel to the current-sheet current and seeds magnetic structures that drive its rapid breakup and dissipation; when injected before the sheet forms, the same-polarity beam field piles up against the inflowing plume fields and suppresses sheet formation.

\subsection*{Simulation results}

\begin{figure}
\centering
    \includegraphics[width=\linewidth]{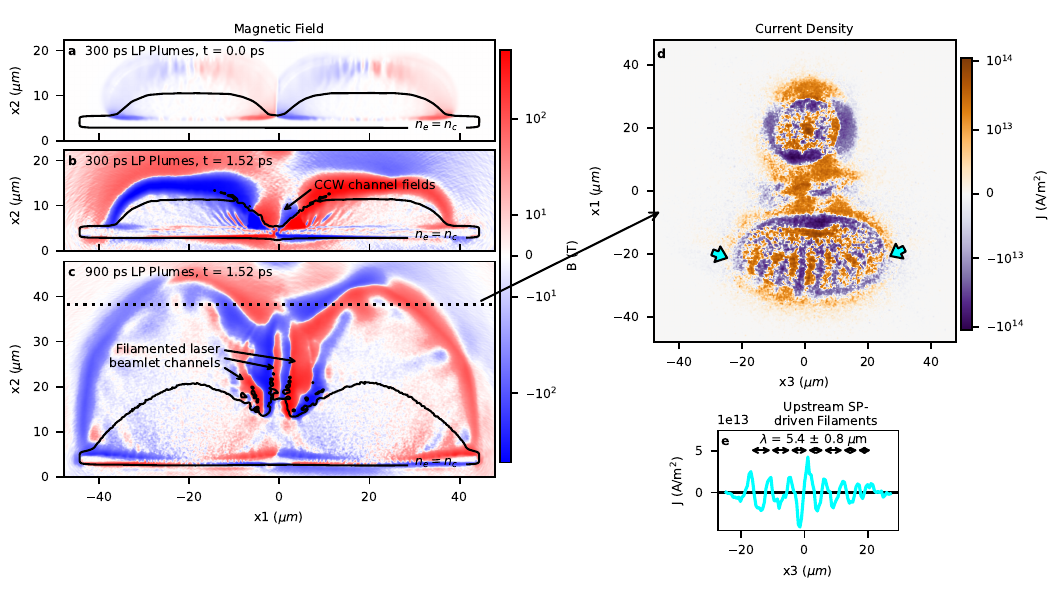}
    \caption{ 
     \textbf{3D particle-in-cell simulation of short-pulse laser-driven fields.}
     Panels \textbf{a}-\textbf{c} show in color the out-of-plane magnetic field in the plane through the center of the LP-driven plumes.
      The black line is a contour of $n_e=n_c$, the critical electron plasma density for the relativistic short-pulse laser.
      Panel \textbf{a} shows the initial condition before the SP interaction, and panels \textbf{b}-\textbf{c} show 1.5 ps after the beginning of the simulation.
      \textbf{a}-\textbf{b} are of the simulation with the $t=t_\text{LP}+300\text{ ps}$ initial condition, and \textbf{c} is with the $t=t_\text{LP}+900\text{ ps}$ initial conditions.
      By $t=\SI{1.5}{\ps}$ in both cases, the scattered laser light has left the simulation box. 
      Panel \textbf{d} shows the longitudinal ($x_2$-direction) current density in a plane capturing the upstream hot-electron-generated magnetic fields, indicated by the dotted line in \textbf{c}.
      A lineout of the current modulations between the arrows in \textbf{d} is shown in panel e in panel \textbf{e}.
      The PIC simulation was initialized from data from the GORGON XMHD simulation, spatially downscaled 25 times.
      }
    \label{fig:e05-l05_b3}
\end{figure}

%\section*{Discussion}

An end-to-end self-consistent 3D model of the laser-plasma interaction was run using a hybrid approach. This was done using the 3D extended-magnetohydrodynamics (XMHD) code GORGON to model the long-pulse irradiation of two colliding plasma plumes and the 3D particle-in-cell (PIC) code OSIRIS to model the third short-pulse relativistic-intensity laser perturbation to the plumes. The output of the XMHD simulation was used as the initial conditions for the PIC simulation (see Methods).
To our knowledge, this is among the first demonstrations of using a 3D fluid model to initialize a 3D PIC simulation \cite{campbellFormationCollisionlessShocks2024}.
The higher resolution requirements of PIC as compared to XMHD are such that the model of the millimeter-scale plasma in 3D had to be spatially downscaled by a factor of 25 in each dimension in order to be computationally feasible.

A 3D numerical model was used to instantiate the LP plasma plumes, therefore the 3D PIC simulation of the SP interaction captures the geometry dependence of the laser-driven electron current generation, reproducing key features from our experimental data analysis.
If the SP arrives before the LP current sheet formation, then it has two distinct magnetic field regions: one comprises the fields generated by radially accelerated electrons, with clockwise polarity, which spreads over the top of the plumes, and the other is the counterclockwise fields of the electron acceleration channel between the two plumes (Fig.\ \ref{fig:e05-l05_b3}\textbf{b}).
The channel field persists after the SP duration is over and is the most likely candidate for the central fields we measure in the experiment that persist 600 ps after the SP (Fig.\ \ref{fig:doryboard} \textbf{f}).
When the SP arrives after the current sheet has formed, the plasma density scale length is also much longer, resulting in elongated channels and reduced Biermann battery fields (Fig.\ \ref{fig:e05-l05_b3}\textbf{c}).
The longer density scale length also results in filamentation of the laser beam into several beamlets, with magnetic channels forming in each beamlet, causing multiple flux ropes to be injected in and around the current sheet.
Electrons with a radially accelerated component create stripes of alternating longitudinal current that penetrate into the top portions of the plumes (Fig.\ \ref{fig:e05-l05_b3}\textbf{d}-\textbf{e}) with a wavelength on the order of the electron inertial wavelength (Supplementary Sec.\ \ref{app:params}), emphasizing the role of particle kinetic instabilities on the upstream magnetic field modification.

\section*{Conclusion}

We have shown that a single relativistic laser pulse can actively control magnetic reconnection in a high-energy-density plasma, either accelerating the breakup and dissipation of an established current sheet by seeding current-sheet instabilities, or suppressing it by building magnetic pressure ahead of sheet formation (Fig.\ \ref{fig:setup}\textbf{a}). In both regimes the laser injects an electron beam whose current is parallel to the Biermann-generated current at the plume collision interface, and the timing of this injection relative to current-sheet formation sets whether the sheet and the upstream magnetic field are driven into rapid, current-filamentation-driven dissipation or reconnection is inhibited altogether. 

Proton deflectometry resolved the two-dimensional field structure with picosecond resolution, and reduced-scale 3D particle-in-cell simulations reproduced the beam-driven field topology and revealed how laser filamentation seeds multiple field columns. Building on this, future experiments could recover the full three-dimensional field (for example, through tomographic proton deflectometry or optical Faraday-rotation probes \cite{griff-mcmahonStructureSelfgeneratedMagnetic2026}) to independently resolve the physics of the different density regions.
 
Because the sign of the effect is set by the polarity of the injected field, driving the electron beam in the opposite direction should reverse it, extending the range of achievable control.
Taken together, these results turn magnetic reconnection into a controllable, tunable laboratory system, enabling on-demand tests of reconnection and plasmoid dynamics, offering a route to steering magnetic-energy release in nuclear fusion plasmas, and broadening the range of conditions accessible to laboratory astrophysics.

\section*{Methods}
\subsection*{Experiment: Laser and Target Parameters}

For the experiment, conducted at the OMEGA-EP laser facility, 
two long-pulse beams (LPs) drove plasma plumes side-by-side which collided, and a high-intensity short-pulse (SP) beam was fired at the plume interface (Fig.\ \ref{fig:setup}).
The experiment utilized three laser pulses for the main interaction on a \SI{50}{\micro\meter}-thick plastic (CH) target and one laser pulse to generate the proton beam (Fig. \ref{fig:setup} \textbf{c}).
The SP on the main target was a 10 ps (FWHM), 1053-nm (IR) laser pulse with 100 J of energy transmitted through an f/6 apodizer, focused to a \SI{25}{\micro\meter} spot 
with peak intensity $1\times10^{19}\text{ W/cm}^2$ and 
$R_{80}$ average intensity $1.6\times10^{18}\text{ W/cm}^2$ ($R_{80}$ is the radius containing 80\% of the laser energy).
It was incident at an angle of 45\textdegree\ to target normal.
Two 351-nm (UV) LP laser pulses were each 2.5 ns long with intensities on the order of $1\times 10^{14} \text{W/cm}^2$. The top UV pulse deposited about 1400 J in a spot of diameter
\SI{800}{\micro\meter}, incident to the target at an angle of 31\textdegree\ to target normal,
while the bottom UV LP had an incident angle of 63\textdegree\ with respect to normal, which elongated the on-target focal spot in one direction. 
In order to have its on-target intensity approximately match that of the top focal spot, its pulse energy was set to 2600 J. 
In order to have the temperature gradients approximately equal and opposite between the two focal spots, the UV foci were arranged such that the top, circular spot was placed along the non-elongated axis of the bottom, elongated spot (Fig.\ \ref{fig:setup} \textbf{c}). 
This was done because the magnetic fields generated by the Biermann battery effect \cite{campbellMeasuringMagneticFlux2022} are proportional to the cross product of the electron temperature gradient with the gradient of the logarithm of the electron density.
The center-to-center separation of the LP focal spots was \SI{1000}{\micro\meter} and the SP hit the target directly between the plasma plumes created by the LP lasers. Three different timings of the SP relative to the LPs were used: $t=t_\text{LP}+300$ ps, $t=t_\text{LP}+500$ ps, and $t=t_\text{LP}+900$ ps, where $t_\text{LP}$ is the time of the onset of the LPs on the target.
The later the SP arrives, the longer the plasma scale length and the greater the magnetic field strength in the plumes.

\subsection*{Experiment Diagnostic: Proton Deflectometry (TNSA$\rightarrow$RCF)}
The main diagnostic was proton deflectometry of the magnetic fields \cite{willingaleProtonDeflectometryMagnetic2010}. A proton beam was generated via 
another IR SP striking a copper target to accelerate protons via the target normal sheath acceleration (TNSA) mechanism \cite{clarkMeasurementsEnergeticProton2000,krushelnickEnergeticProtonProduction2000,maksimchukForwardIonAcceleration2000,clarkEnergeticHeavyIonProton2000,snavelyIntenseHighEnergyProton2000,wilksEnergeticProtonGeneration2001,mckennaCharacterizationProtonHeavier2004,fuchsLaserdrivenProtonScaling2006,macchiIonAccelerationSuperintense2013}.
The resulting proton beam first passed through a \SI{5}{\micro\meter} tantalum shield to shield the proton target from radiation. The proton beam then passed through the main interaction target, where the plasma magnetic and electric fields spatially modulated the beam before the proton beam arrived at the stack of radiochromic film which served as the proton detector. The deflection of charged particles to probe electromagnetic fields \cite{kuglandInvitedArticleRelation2012,bottProtonImagingStochastic2017,daviesEvaluationDirectInversion2022,daviesQuantitativeProtonRadiography2023,schaefferProtonImagingHighenergydensity2023} is an excellent diagnostic because it provides 2-D (z-integrated) information with time resolution on the order of picoseconds.
The TNSA proton source creates a wide spectrum of proton energies, resulting in the probe beam spreading out in time-of-flight, allowing a single shot to be probed at multiple points in time. The different points in time can be recorded in different layers of film in the detector stack because of the Bragg peak of the proton energy deposition which causes the faster protons to deposit most of their energy deeper into the detector stack \cite{knollRadiationDetectionMeasurement2010}.
This technique allows for many proton images with time resolution on the order of a few ps all within a single shot, limited by the blurring from the high-energy protons depositing a portion of their energy on the films closest to the source of the proton beam \cite{zylstraUsingHighintensityLasergenerated2012}.

The proton-generating IR pulse had 300 J (for some shots, 150 J) at best temporal compression, which is typically about 700 fs, at best focus, which had 80\% of the laser energy within a radius of less than \SI{18}{\micro\meter}, with peak intensity of about $3\times 10^{20}\text{ W/cm}^2$. The copper ``proton target'' was \SI{50}{\micro\meter} thick, and the tantalum shield was 1.25 mm behind the copper target. The separation between the copper target and main interaction target was 8 mm, and the separation between the main interaction target and the proton detector stack was about 80 mm, resulting in a magnification of about $M=11$. The proton detector stack consisted of 25 layers of GafChromic HD-V2 film \cite{binAbsoluteCalibrationGafChromic2019} interspersed between aluminum filters of thickness varying from \SI{100}{\micro\meter} to \SI{3000}{\micro\meter}. The stack was designed to capture protons of energy 9.5 MeV to 40.7 MeV, resulting in a time-of-flight range of 95 ps being measured for each shot.

In Fig.\ \ref{fig:doryboard}, the main proton energy is
    17.8 MeV for \textbf{a-b}; % (H06),
    22.6 MeV for \textbf{c-e} and \textbf{g}; %(H09), 
    and 30.7 MeV for \textbf{f}. %(H18)

\subsubsection*{Processing RCF Data: Optical Density to Dose}
We interpreted the RCF images using the following steps: We calibrated our scanner (EPSON V850 Pro) with an 11-step ND filter (OD 0.04 to 2.0, from Edmund Optics) to convert pixel value to optical density (OD). Then we used the calibration from Bin et al.\ \cite{binAbsoluteCalibrationGafChromic2019} to convert OD to dose for GafChromic HD-V2 film. Even though an accurate dose measurement would have needed the calibration factors for our particular batch of film, using the calibration from Bin et al. gives a better understanding of the proton distribution than does optical density alone.

The final critical step in analyzing proton radiochromic film
data is to obtain an estimate of $I_0(x,y)$; i.e., what was the proton beam
intensity as a function of x and y before it passed through the
plasma?
There is enough shot-to-shot fluctuation in the proton beam profile to the extent that using a separate null shot with no target plasma for $I_0$ would cause significant analysis errors.
The most widely used method to estimate $I_0$ is to
apply a 2D low-pass filter to the digitized RCF data, i.e., Fourier
filtering.
However, the resulting magnitude and topology of the reconstructed magnetic field can vary significantly depending on the choice of Fourier filter. 
A practical approach is to scan through several choices of Fourier filter and 
examine the variation between the outputs' magnetic topologies.

\subsection*{Simulation - Rad-MHD setup}

First, to simulate the creation of the plasma plumes by the long-pulse lasers, the rad-hydro code GORGON \cite{walshExtendedMagnetohydrodynamicEffects2018a,walshExtendedmagnetohydrodynamicsUnderdensePlasmas2020} was used in 3D.
Then the plasma density, ion and electron temperatures, fluid velocity, and magnetic fields calculated from the GORGON simulation were used as the initial conditions for a fully-3D particle-in-cell (PIC) simulation using the OSIRIS 4.0 framework \cite{fonsecaOSIRISThreeDimensionalFully2002}, simulating the arrival of the SP onto the target.
Relativistic laser-plasma interactions ($I\geq 10^{18}\,\unit{\watt\per\square\cm}$) are not well modeled by XMHD simulations because the non-Maxwellian velocity distribution of electrons has an immense influence on the results of the experiment. 

The GORGON simulation had UV LP lasers set up just like in the experiment, with the same incidence angles and intensity profiles, and at the same spatial scale.
The simulation parameters are summarized in Table \ref{tab:gorgon}.
%%Two beams, both with SG8-0750 phase plate, 2.5 ns duration.
%%Beam 3: 2600 J, 63-degree angle of incidence.
%%Beam 4: 1400 J, 31-degree angle of incidence.
%%Center-center separation of 1.0 mm. 
%%The geometry of the lasers are as described in Figs. \ref{fig:incident_side} and \ref{fig:incident_top-down}
%
\begin{table}
\centering
\begin{tabular}{cc}
\hline  \hline 
Parameter & Value \\
\hline
$N_x$, $N_y$, $N_z$ & 360, 360, 240 \\
$L_x$, $L_y$, $L_z$ & \SI{4000}{\micro\meter}, \SI{4000}{\micro\meter}, \SI{1200}{\micro\meter} \\ 
$d_x$, $d_y$, $d_z$ & \SI{11.1}{\micro\meter}, \SI{11.1}{\micro\meter}, \SI{5}{\micro\meter} \\
Beam 3 Incidence Angle & 63$^\circ$ \\
Beam 3 Pulse Energy & \SI{2600}{\joule} \\
Beam 4 Incidence Angle & 31$^\circ$ \\
Beam 4 Pulse Energy & \SI{1400}{\joule} \\
\hline 
\hline
\end{tabular}
\caption{\label{tab:gorgon} \textbf{Parameters for the GORGON simulation of long-pulse laser-driven plume development.}
Beam 3 and Beam 4 are the names of the corresponding beamlines of OMEGA EP.}
\end{table}

\subsection*{Simulation - PIC setup}

The OSIRIS simulations were in full 3D in Cartesian geometry with open electromagnetic and particle boundary conditions,
run on 64,000 AMD EPYC 7742 (Rome) cores.
The simulation geometry was spatially downscaled 25 times in
order to make it computationally feasible, and the SP was incident
normal to the target, rather than at 45\textdegree.
The porting of the GORGON data to the initial conditions of the OSIRIS simulation was done using 
a modified version 
of OSIRIS.
The OSIRIS simulation parameters are summarized in Table \ref{tab:osiris}.
Simulations with a little less than half the resolution in each dimension showed similar results but with extensive numerical diffusion, indicating that the full resolution simulation is converging toward a numerically stable result.
\begin{table}
\centering
\begin{tabular}{cc}
  \hline \hline 
Parameter & Value \\
\hline$n_{\text{cores},x3}$ & 256, 1, 250 \\
$N_{x1}$, $N_{x2}$, $N_{x3}$ & 1800, 900, 1800 \\
$L_{x1}$, $L_{x2}$, $L_{x3}$ & 600 $c/\omega_L$, 300 $c/\omega_L$, 600 $c/\omega_L$ \\ 
$d_x$, $d_y$, $d_z$ & $\frac{1}{3}\,c/\omega_L$, $\frac{1}{3}\,c/\omega_L$, $\frac{1}{3}\,c/\omega_L$ \\
$d_t$ & $0.1905/\omega_L$ \\
$M_i/m_e$ & 3672.3 \\
ppc$_i$, ppc$_e$, & 27, 64 \\
$a_0$ & 2.2 \\
Laser propagation direction & $x_2$ \\
Laser polarization angle & 50$^\circ$ \\ 
$t_\text{env,duration}$ & $859.23\omega_L^{-1}$ (FWHM = 400 fs) \\
$t_\text{env,range}$ & $1843.9\omega_L^{-1}$\\
$\text{per}_\text{focus}$  & 33.3 $c/\omega_L$\\
$\text{per}_\text{fwhm}$ & 19.74 $c/\omega_L$ \\
\hline \hline
\end{tabular}
\caption{\label{tab:osiris}\textbf{Parameters for the OSIRIS simulation.}
  Lengths and times are normalized to the laser frequency $\omega_L$ and the speed of light in vacuum $c$.
  $L_x$ the length of the simulation box,
  $d_x$ the spatial resolution of the grid,
  $M_i$ and $m_e$ the ion and electron masses, respectively,
  ppc the number of initialized particles per cell for each species,
  $a_0$ the dimensionless vector potential corresponding to the focused laser intensity,
  $t_\text{env,duration}$ specifies the $\sigma_t$ of the gaussian laser pulse envelope,
  $t_\text{env,range}$ defines the numerical start and stop of the laser in the simulation (in this case, defined such that the laser starts and stops at 1\% of its peak intensity),
  $\text{per}_\text{focus}$ defines the $x_2$ position of the laser focus,
  and $\text{per}_\text{fwhm}$ is the full-width at half-maximum size of the laser focal spot.
  }
\end{table}

\section*{Data availability}

The data that support the findings of this study are available from the corresponding author upon reasonable request.

\section*{Code availability}

The 3D PIC simulations were run in OSIRIS, which is open source. 
The modified open-source code used here can be provided upon reasonable request.
The 3D XMHD simulations were run in GORGON, which is not open source. 

%\nocite{*}
\bibliography{bigbibtex}
% \bibliography{da_reference}% Produces the bibliography via BibTeX.

\section*{Acknowledgements}
  The experiment was conducted at the Omega Laser Facility with beam time provided through the National Laser Users' Facility (NLUF) under the auspices of the U.S. DOE/NNSA by the University of Rochester's Laboratory for Laser Energetics under Contract DE-NA0004144. 
    
C.D. is supported by DOE grant DE-SC0024639, NASA grant 80NSSC25K0050, NSF grant AGS-2301338, the Alfred P. Sloan Research Fellowship, and the IBM Einstein Fellow Fund at the Institute for Advanced Study, Princeton.   
    
 K.G.M. discloses support for the research of this work from the DOE Office of Fusion Energy Sciences under Award Number DE-SC0021057 and the DOE National Nuclear Security Administration (NNSA) through the University of Rochester's 'National Inertial Confinement Fusion Program' under Award Number DE-NA0004144.
  
We thank the OMEGA-EP team for skillful execution of the experiment, for help with planning the diagnostics, and for help designing the experiments to maximize the potential of the facility.

We thank the OSIRIS Consortium for making the OSIRIS code publicly available.

We acknowledge high-performance computing support for the 3D PIC simulations from the National Energy Research Scientific Computing Center, a DOE Office of Science user facility; from the NASA High-End Computing (HEC) Program through the NASA Advanced Supercomputing (NAS) Division at Ames Research Center; and from the Derecho system (\url{doi:10.5065/qx9a-pg09}) provided by the NSF National Center for Atmospheric Research (NCAR), sponsored by the National Science Foundation.

The XMHD simulation work was performed under the auspices of the U.S.
Department of Energy by Lawrence Livermore National Laboratory
under Contract No. DE-AC52-07NA27344.
This document was prepared as an account of work sponsored
by an agency of the United States government. Neither the United
States government nor Lawrence Livermore National Security, LLC,
nor any of their employees makes any warranty, expressed or
implied, or assumes any legal liability or responsibility for the
accuracy, completeness, or usefulness of any information,
apparatus, product, or process disclosed, or represents that its use
would not infringe privately owned rights. Reference herein to any
specific commercial product, process, or service by trade name,
trademark, manufacturer, or otherwise does not necessarily
constitute or imply its endorsement, recommendation, or favoring
by the United States government or Lawrence Livermore National
Security, LLC. The views and opinions of authors expressed herein
do not necessarily state or reflect those of the United States
government or Lawrence Livermore National Security, LLC, and
shall not be used for advertising or product endorsement purposes.

\section*{Author Contributions}
J. L. Latham led the writing of the manuscript, led the experiments at OMEGA-EP, implemented the modified version of OSIRIS
to port GORGON data into the OSIRIS initial conditions, led the experimental data analysis and simulation data presentation.  

B. K. Russell contributed to writing the manuscript, helped plan the experiment, helped with the PIC simulation, and helped guide the direction of the work.

C. Dong led the 3D PIC simulation effort, contributed to writing the manuscript, and helped guide the direction of the work.

C. A. Walsh performed the XMHD simulations and edited the manuscript.

K. G. Miller wrote the modified version of OSIRIS used to port the MHD data to the PIC simulation initialization routine and edited the manuscript. 

P. T. Campbell helped plan the experiment, helped with data analysis and interpretation of the results, and edited the manuscript.

L. Willingale helped plan the experiment, contributed to interpreting the results, and edited the manuscript.

P. Nilson helped develop the general experimental concept, helped interpret the results, and reviewed the manuscript.

K. Krushelnick developed the experimental concept, helped interpret the experimental results, guided the direction of the work, and edited the manuscript.

\section*{Competing Interests}
The authors declare no competing interests.

\section*{Additional Information}

\noindent Supplementary Information is available for this paper.

\noindent Correspondence and requests for materials should be addressed to J. L. Latham (joshla@umich.edu).

\clearpage

\subsection*{Supplementary Information}

% Number the supplementary sections as S1, S2, ... (they were rendering as 0.0.1)
\setcounter{section}{0}
\renewcommand{\thesection}{S\arabic{section}}

\section{Calculation of Parameters}
\label{app:params}
The GORGON simulation of the long-pulse (LP) plume development, with \SI{1e14}{\watt\per\square\cm} UV lasers, resulted in an electron temperature of $T_e\sim 10^3\text{ eV}$.
The simulation gave the same temperature for ions, $T_i\sim 1\text{ keV}$.
This value of temperature agrees with other literature \cite{hironakaIdentificationStimulatedRaman2023} reporting that for a laser at $10^{15}\text{ W/cm}^2$ the plasma is 1 keV to 3 keV.
Given the lower intensity of our lasers compared to Ref.\ \cite{hironakaIdentificationStimulatedRaman2023}, and given the possibility that the simulation overestimates the laser absorption, we estimate a possible temperature range from \SI{100}{\eV} to \SI{1}{\keV}.

The magnetic field strength $B$ in the high-field regions
(in a $\sim$0.2 mm thick rim around each plasma plume)
was 100 T according to the GORGON simulation,
and 50 T according to Ref.\ \cite{rosenbergLaboratoryStudyAsymmetric2015} (which had 500 J, 1 ns pulse in \SI{800}{\um} spots and an intensity comparable to those in the present experiment).
According to our calculations, we measure 140 T.
The deflectometry image inversion of the long-pulses only after about 900 ps of irradiation (Fig.\ \ref{fig:37101bfieldinversion}) 
calculated an integrated magnetic field of about $18 \text{T}\cdot\text{mm}$;
based on the angular filter refractometry measurements (AFR) of a laser shot with half the intensity, 
we estimate a height of 0.13 mm $\pm$ 0.05 mm.
So, 100 T $\pm$ 50 T is our range for magnetic field strength in the highly magnetized region.
In the lower density corona the GORGON simulation reports a magnetic field strength of about 1 T. 

For electron density $n_e$ we measure $\sim 10^{21}\text{ cm}^{-3}$ from the AFR measurements and the GORGON simulation. This is about the critical density for the SP (wavelength of \SI{1.05}{\micro\meter}).
Some important magnetic interactions also occur at slightly above the critical density, as well as down to 0.05 times the critical density (about 0.006 times the critical density for the $3\omega$ UV lasers). 

For mass density $\rho$ we have 0.1 to 10 kg/m$^3$ from the GORGON simulation, depending on the distance away from the target.
At the critical surface for fully ionized CH, the mass density is $3.1\text{ kg/m}^3$.
Given the range of possible electron densities $0.01n_c$ to $n_c$, this gives a range of $\rho$ from \SI{0.03}{\kg\per\cubic\m} to \SI{0.3}{\kg\per\cubic\m}. 

The magnetic field and mass density together give the Alfv\'en speed, $v_A = B/\sqrt{\mu_0\rho}$.
Using the opposite extremes of magnetic field, 50 T to 150 T,
and mass density, 10 kg/m$^3$, to \SI{0.03}{\kg\per\cubic\m},
this results in a range of Alfv\'en speeds from
$\SI{25}{\um\per\ns}< v_A < \SI{760}{\um\per\ns}$.

The Spitzer resistivity \cite{cohenElectricalConductivityIonized1950,spitzerTransportPhenomenaCompletely1953} in SI units is
\begin{equation}
  \eta_\text{Spitz}=1.15\times10^{-14}\,Z\ln\Lambda\, T^{-3/2}\text{ sec}
  \times \left(\frac{1}{9}\times 10^{-9}\text{ sec}/(\Omega\cdot\text{m})\right)^{-1}.
\end{equation}
Using $Z=3.5$, $\text{ln}\Lambda =15$,
and $T$ from \SI{100}{\eV} to $1000\text{ eV}$,
we get $\eta_\text{Spitz}$ from  \SI{5.4e-6}{\ohm\m} to $1.7\times 10^{-7}\text{ }\Omega\cdot\text{m}$. 

Putting these together, we calculate the Lundquist number, $S = \mu_0 v_A L/\eta$, where $L$ is the length scale.
The maximum length scale is \SI{1}{\mm}, the diameter of the plumes and approximate length of the current sheet.
The minimum length scale is \SI{100}{\um}, the plasmoid width; plasmoids reduce the effective length scale and accelerate reconnection.
Again, using the extremes of the Alfv\'en speed $v_A$, $L$, and $\eta$ to estimate the extremes of the Lundquist number in our experiment,
our range is $0.6 < S < 5.5\times 10^4$.
So in the relatively cold, unmagnetized, dense regions of the plasma in our experiment, resistive effects will dominate over magnetic perturbations.
But, in the highest-magnetic-field, hottest, and most tenuous of the plasmas in the long-pulse plumes in our experiment, the Lundquist number is estimated to be in the bottom of the range applicable to plasmoid formation theory. 

Another important parameter for reconnection studies is plasma $\beta$, the ratio of plasma pressure to magnetic pressure:
\begin{equation}
  \beta = \frac{nT}{B^2/2\mu_0}
\end{equation}
For the minimum $\beta$ in the LP plumes,
with $n_e\sim\SI{1e19}{\per\cubic\cm}$, $T_e\sim\SI{100}{\eV}$, and $B\sim\SI{150}{\tesla}$, we find $\beta=1.8\times 10^{-2}$. 
For the opposite extreme, $n_e\sim10^{21}\text{ cm}^{-3}$, $T_e\sim1\text{ keV}$, and $B\sim50\text{ T}$, we get $\beta=160$.
Thus, in the most tenuous, most magnetized parts of the LP plumes, magnetic effects dominate over pressure effects by a factor of $\beta^{-1}\sim 50$,
but in the denser, hotter, less magnetized parts of the plasma, plasma thermal pressure effects dominate over magnetic effects by a factor of up to $\beta\sim 160$. 
This range of $\beta$ puts the LP experiment in an interesting regime where simultaneously there is thermal-pressure-dominated plasma in one point in space and magnetically-dominated plasma in another point in space. 

Timescales: The Alfv\'en time is $\tau_A=L/v_A$. To obtain a lower limit on the Alfv\'en time, we consider the larger possible Alfv\'en speed. With length scale $L$ of 1 mm and Alfv\'en speed $v_A$ of \SI{225}{\micro\meter/\nano\second}, $\tau_A$ is 4.4 ns. 
The resistive timescale is $\tau_R=\mu_0 L^2/\eta$, where $\mu_0$ is the vacuum permeability. For $\eta$ we will use $\eta_\text{Spitz}$. Using $L=1\text{ mm}$ and the hotter $\eta_\text{Spitz}$ above we have $\tau_R$ of \SI{7.3}{\micro\second}. Note that if we consider a smaller length scale our value of $\tau_R$ will be smaller. For example, the width of the current sheet is on the order of \SI{100}{\micro\meter}, and using this value for $L$ the value of $\tau_R$ drops to 73 ns. 
The Sweet--Parker reconnection timescale is the geometric mean of the Alfv\'en time and the resistive time: $\tau_\text{SP} = (\tau_A\tau_R)^{1/2}$. Given our range of possible resistive timescales this gives us a Sweet-Parker time of somewhere between 18 ns and 180 ns. All of the above timescales are beyond the duration of the laser experiment, which only went up to 1.9 ns. Compare this to the analysis in Ref.\ \cite{nilsonMagneticReconnectionPlasma2006}.

Also let us compare the Alfv\'en speed to the plasma flow speed. 
Using the diameter of the top (non-elongated) plume in the proton deflectometry images, we measure that the LP-driven plumes expand at 50 um/ns from t=300 ps to t=500 ps and at 240 um/ns from t=500 ps to t=900 ps. So from early on the plume expansion is super-Alfv\'enic at the critical surface, and at later times the plume expansion is super-Alfv\'enic compared to all except the most extreme upper estimate of $v_A$ for the more tenuous parts of the plasma. 

The sound speed is $C_s = (\gamma Z k T_e/m_i)^{1/2}$.
Using $\gamma=3$, $T_e = \SI{1000}{\eV}$, and $m_i = (m_\text{Carbon} + m_p)/2$, we get $C_s \sim \SI{400}{\micro\meter/\nano\second}$. This is a bit faster than the plume expansion that we measure.

Length scales: 
At the critical density for the \SI{1.05}{\um} laser, which is $n_\text{cr}=\SI{1.01e27}{\per\cubic\meter}$, the ion inertial length $c/\omega_{p,i}$ for hydrogen ions is \SI{7.2}{\um}.  
A full wavelength given this length scale is $2\pi$ times this value, which is \SI{45}{\um}. 
At a density of $0.1 n_\text{cr}$, this wavelength is \SI{142}{\um}. 
The electron inertial length $c/\omega_{pe}$ at the critical density for the infrared laser is equal to the wavelength of the laser divided by $2\pi$, $\SI{0.17}{\um}$.  
At $0.1n_\text{cr}$, the electron inertial length is \SI{0.52}{\um}, so the wavelength associated with the electron inertial length is \SI{3.3}{\um}.

\section{Magnetic Field Inversion}

\begin{figure}
  \centering
  \includegraphics[width=\textwidth]{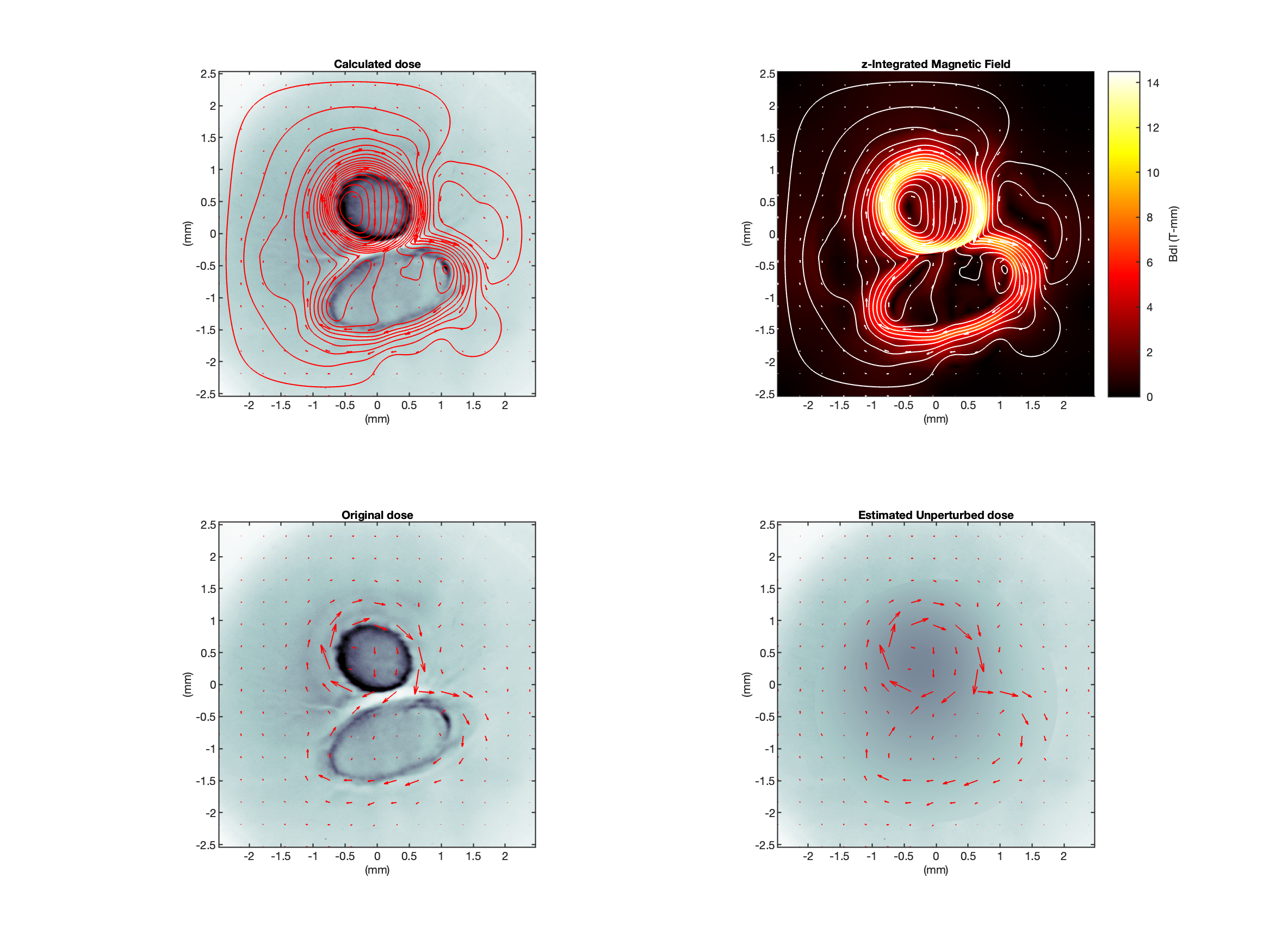}
  \caption{\textbf{Numerical inversion of deflectometry image of LP-only reconnection.}
  This estimates the magnetic
    field of deflectometry image in Fig.\ \ref{fig:doryboard}\textbf{b}, which
    is the long-pulse plumes unperturbed at
    $t=t_\text{LP} + 900\text{ ps}$. Panel \textbf{d} shows the $I_0$
    estimated unperturbed proton dose profile used for the
    calculation. The Fourier filter used was
    $G(\bm{k}) = e^{-|k|^p/(2*1/\lambda_\text{filt})^p}$, with $p=3$
    and $\lambda_\text{filt} = 3.0\text{ mm}$. Note that around the
    edge of the deflectometry image, $I_0$ was set to exactly equal $I$ beyond
    a radius 1.9 mm from the center of the deflectometry image. Panel
    \textbf{b} shows the calculated integrated-magnetic-field strength
    in the heatmap, direction in the quiver plot, and flux
    contours. Panel \textbf{a} shows the re-calculated proton dose
    profile given the unperturbed dose of panel \textbf{d} and the
    magnetic field of panel \textbf{b}, along with the flux contours
    and magnetic field vector arrows. Panel \textbf{c} shows the
    original proton dose profile, with the calculated magnetic field
    vectors.
  }
  \label{fig:37101bfieldinversion}
\end{figure}

\begin{figure}
  \centering
  \includegraphics[width=\textwidth]{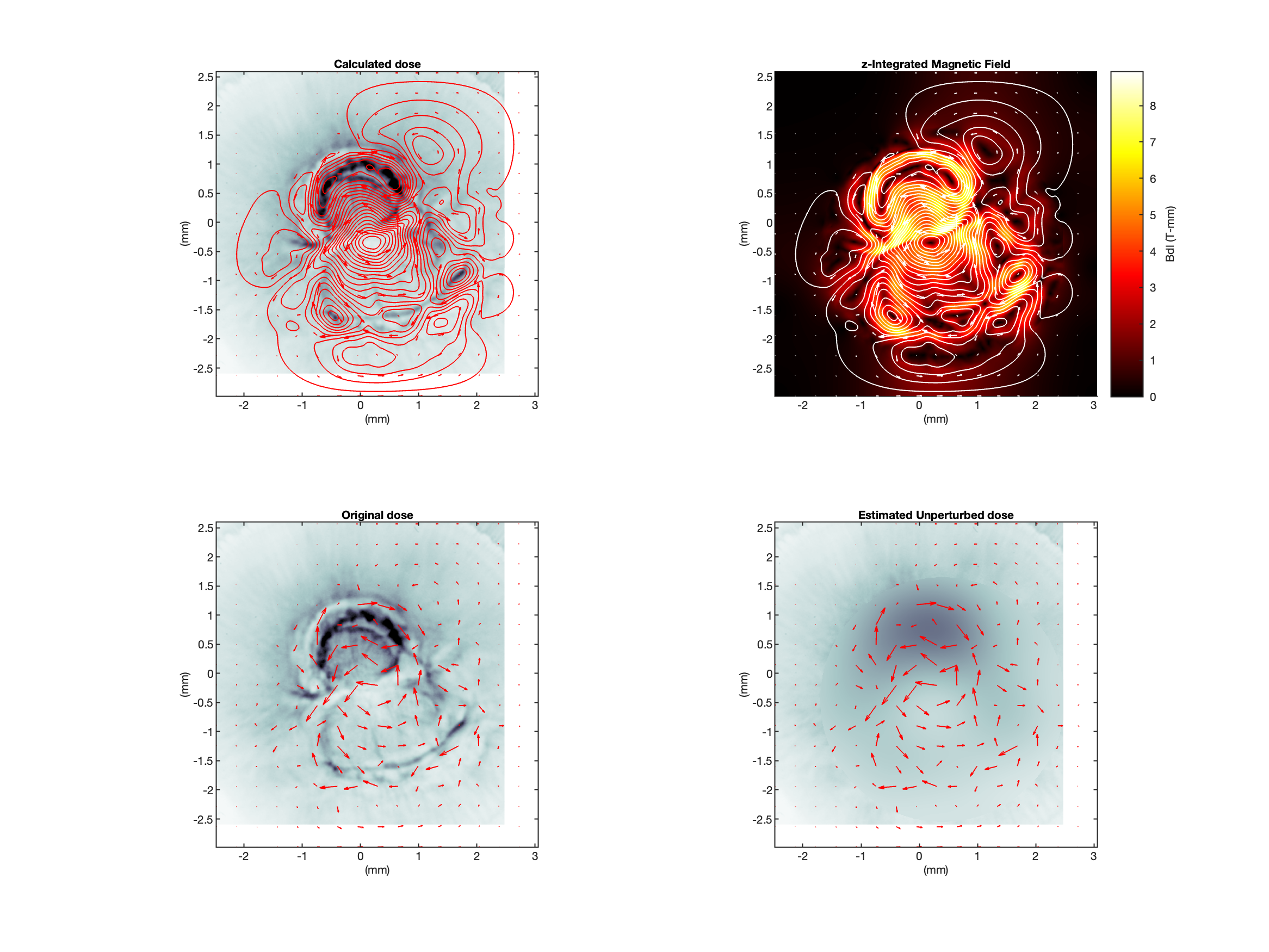}
  \caption{\textbf{Numerical inversion of deflectometry image of SP perturbation to LP-driven reconnection.} 
  This estimates the magnetic
    field of deflectometry image in Fig.\ \ref{fig:doryboard}\textbf{d}, which
    is the long-pulse plumes about 52 ps ($\pm 25$ ps) after the SP perturbation, at 
    $t=t_\text{LP} + 940\text{ ps}$. Panels \textbf{a}, \textbf{b},
    \textbf{c}, and \textbf{d} correspond to those of
    Fig.\ \ref{fig:37101bfieldinversion}, except that for panel \textbf{d}, $p=3$ and $\lambda_\text{filt}=2.0\text{ mm}$.}
  \label{fig:37102bfieldinversion}
\end{figure}

The proton deflections $\alpha_x$ and $\alpha_y$ resulting from the electric and magnetic fields in the paraxial approximation follow the form
\begin{multline} \label{eq:alphax}
  \alpha_x = \frac{v_x}{v_p} = \frac{1}{v_p}\,\,
  \frac{q_p}{m_p}\int dt (E_x(\bm{x}) + v_pB_y(\bm{x}))
  \\\approx
  \frac{1}{v_p}\,\,\frac{q_p}{m_p}\int \frac{dz}{v_p}
  (E_x(x,y,z) + v_pB_y(x,y,z)), 
\end{multline}
where $v_p$ is the proton velocity, $\bm{E}$ is the electric field, $\bm{B}$ is the magnetic field, $q_p$ is the proton charge, $m_p$ is the proton mass, and the proton is propagating mostly in the $\hat{z}$ direction.  
Only the $x$ component of the deflection is shown for simplicity.
$\alpha_x$ and $\alpha_y$ can be written as the $x$ and $y$ derivatives of a potential function $\Phi$, where 
\begin{equation} 
  \frac{I_0}{I}
  =
  1
  +
  \frac{\partial^2 \Phi}{\partial x^2} + \frac{\partial^2 \Phi}{\partial y^2}
  +
  \frac{\partial^2 \Phi}{\partial x^2}\frac{\partial^2 \Phi}{\partial y^2}
  -
  \frac{\partial^2 \Phi}{\partial y\partial x}
  \frac{\partial^2 \Phi}{\partial x\partial y}.
\end{equation}
Here $I_0$ is the $x$ and $y$ distribution of protons before passing through the main target and $I$ is the distribution after the deflections resulting from the interaction of the proton beam with the main target.
This is a Monge--Amp\`ere equation. Given $I_0$ and $I$, one can
solve for $\Phi$ with an iterative Monge--Amp\`ere solver such as
PROBLEM \cite{tzeferacosPROBLEMSolverPROtonimaged2018}. % (\url{https://github.com/flash-center/PROBLEM/tree/master}).
However, one can also solve for $\Phi$ using an iterative Poisson
solver, which is much faster computationally but less robust for
highly stochastic fields. The Poisson solver is detailed in
Ref.\ \cite{campbellLaboratoryInvestigationsMagnetic}, based on
the framework set out by
Kugland \textit{et al.} \cite{kuglandInvitedArticleRelation2012}.
The inverted magnetic field profile for the two LP irradiation case at $t_\text{LP}+900\text{ ps}$ is shown in Fig.\ \ref{fig:37101bfieldinversion}. 
The same for the case after the SP arrives at $t_\text{LP}+900\text{ ps}$ is shown in Fig.\ \ref{fig:37102bfieldinversion}.

\section{LP Only Plasmoid Formation}
\label{sec:lponlyplasmoid}

The deflectometry images of the long-pulse-only shots show evidence of plasmoid formation in the plume-interaction current sheet and in the sheets of current on the outer edges of the plumes.
From 825 ps after the start of the long-pulse drive to 972 ps after the long-pulse drive, the number of dark bands in the white proton-depleted image of the interaction current sheet grows from 1 band at 825 ps, 2 bands at 860 ps, and 3 bands at 972 ps (Fig.\ \ref{fig:37102_superlineout}\textbf{d-f}).
If we use the growth in the number of plasmoids as a proxy for the growth rate of the instability, then we have a growth time (inverse of growth rate) $\gamma^{-1}=73\pm\SI{38}{\ps}$.
The wavelength of these dark bands is approximately $\lambda=\SI{112}{\um}$.
This wavelength matches the wavelength of modulations on the bottom side of the elongated plume (Fig.\ \ref{fig:37102_superlineout}\textbf{c}, not annotated).
The outer edge of the elongated plume also has modulations with $\lambda=\SI{160}{\um}$ (Fig.\ \ref{fig:37102_superlineout}\textbf{c}, not annotated),
and the outer edge of the top plume has modulations that range from $\lambda=\SI{120}{\um}$ to $\lambda=\SI{200}{\um}$ (Fig.\ \ref{fig:37102_superlineout}\textbf{a-c}, not annotated).

\end{document}